# A nonlinear and time-dependent visco-elasto-plastic rheology model for studying shock-physics phenomena

*Dirk Elbeshausen[1] and H. Jay Melosh[2*]*

[1]Autodesk GmbH, Försterweg 3, 14482 Potsday, Germany. [2]Purdue University, Earth, Atmospheric, and Planetary Sciences Department, 550 Stadium Mall Drive, West Lafayette, IN 47907-2051, USA

*Corresponding Author

## Abstract
We present a simple and efficient implementation of a viscous creep rheology based on diffusion creep, dislocation creep and the Peierls mechanism in conjunction with an elasto-plastic rheology model into a shock-physics code, the iSALE open-source impact code. Our approach is based on the calculation of an "effective viscosity" which is then used as a reference viscosity for any underlying viscoelastic (or even visco-elasto-plastic) model. Here we use a *Maxwell-model* which best describes stress relaxation and is therefore likely most important for the formation of large meteorite impact basins. While common viscoelastic behavior during mantle convection or other slow geodynamic or geological processes is mostly controlled by diffusion and dislocation creep, we showed that the Peierls mechanism dominates at the large strain rates that typically occur during meteorite impacts. Thus, the resulting visco-elasto-plastic rheology allows implementation of a more realistic mantle behavior in computer simulations, especially for those dealing with large meteorite impacts. The approach shown here opens the way to more faithful simulations of large impact basin formation, especially in elucidating the physics behind the formation of the external fault rings characteristic of large lunar basins.

## 1 Introduction

Viscoelastic behavior of material has been studied intensively over the last decades. Studies range from engineering and industrial applications (e.g. Darabi et al., 2011; Larsen et al., 2009; Morian et al., 2014; Hong, 2011) as well as biological (e.g. Grooman et al., 2012; Rouse Jr., 2004), medical (Gennissson et al., 2010; Streitberger et al., 2011; Teran et al., 2010) and geological objectives (Lange et al., 2014; Remus et al., 2011; Wang et al., 2012; Nield et al., 2014; Theofanous, 2011). The concept of viscoelasticity was first proposed by Maxwell in 1867 (Maxwell, 1867). Later on, progress in the understanding of crystal structures of geological materials resulted in improved mathematical and physical models [Karato, 2010], such as the diffusion creep model [Nabarro, 1948] or a power-law dislocation creep model [Weertman, 1955]. These and other improvements to the description of viscoelastic behavior contributed to a better understanding of (geo)dynamical problems, such as plate tectonics, mantle convection, mountain ridge formation or dynamic earthquake ruptures. Over the last decades, extensive efforts have been made to develop and improve numerical rheology models to further increase the realism of numerical simulations.

It is widely known that the viscous rheologic properties of the mantle material have a significant effect on tectonic processes, such as subduction, folding [e.g. Hobbs et al., 2007], mountain formation [Patton and Watkinson, 2010] and even the growth of fractures and faults [Nguyen et al., 2013]. It has been also established that extremely large meteorite impacts could have induced significant deformation of the Earth's mantle during the Hadean era of heavy meteorite bombardment [Christeson et al., 2009; Ivanov, 2005; Ivanov et al., 2010; Potter et al., 2013]. However, most previous numerical simulations failed to include a proper model for the viscous deformation of mantle material. In contrast to many other geodynamical applications, where both the timescales and the length scales are long (up to millions of years for the formation of mountains), meteorite impact is a very rapid process with relevant timescales ranging from milliseconds to seconds [Melosh, 1989] and strains many times larger than 100%. Thus, it is plausible that small-scale phenomena occurring under high deviatoric stresses, such as changes in the crystal lattice, might become important in describing the macroscopic effect of viscoelasticity during impact processes. Furthermore, meteorite impacts generate shock waves (compressive waves) that propagate with speeds higher than the sound speed in the surrounding material. Irreversible compression and adiabatic decompression of the material occurs, resulting in different types of shock effects (such as vaporization, melting, fracture, collapse or opening of pore spaces), depending on the structure of the shock wave and the subsequent rarefaction wave. Some authors argue that the viscosity of geologic materials might be able to initiate a shock viscosity [Melosh, 2003; Swegle and Grady, 1985; Benson, 1991] that can broaden the shock wave and therefore might change the resulting shock effects. However, lacking of a proper visco-elasto-plastic rheology model for shock-physics codes, this has not been carefully tested. Here we present our implementation of such a rheology into the shock-physics code iSALE.

## 1.1 The iSALE hydrocode

The hydrocode *iSALE* (Elbeshausen et al., 2009; Elbeshausen and Wünnemann, 2011; Wünnemann et al., 2006; Collins et al., 2004; http://www.isale-code.de) is a multi-material shock physics code (historically called "hydrocode"). It solves the Navier-Stokes equations in a compressible fashion on a Cartesian staggered mesh by using finite differences and finite volumes in two and three dimensions. The solver follows the scheme of Hirt et al., 1974 which allows the calculation of flows at nearly arbitrary speeds. The simulations can be performed in an Eulerian or Lagrangian fashion as well as by using an Arbitrary Lagrangian Eulerian (ALE) technique. For calculations of meteorite impacts, however, the large strains would greatly distort a Lagrangian mesh, making the Eulerian approach the most reasonable one. The code additionally consists of a number of different equations of state, such as ANEOS [Thompson and Lauson, 1972] or Tillotson [Tillotson, 1962], allowing computations with a realistic thermodynamic behavior for many geomaterials. In addition different constitutive models are required to account for material damage and failure [Collins et al., 2004; Ivanov et al., 1997], porosity [Wünnemann et al., 2006], dilatancy [Collins, 2015], acoustic fluidization [Wünnemann and Ivanov, 2003], thermal softening [Ohnaka, 1995], or low-density weakening.

The underlying constitutive model, however, is either elasto-plastic or purely viscous (simple linear and time-independent viscosity). A visco-elastic or even visco-elasto-plastic rheology has not been considered yet. The development and implementation of this extension is presented in the subsequent sections.

## 2  The Peierls mechanism – description

The rheology of rocks depends on a large number of constitutive and environmental factors, including mineralogy, fluid content and chemistry, mineral grain size (see e.g. Riedel and Karato, 1997), melt fraction, temperature, pressure, and differential stress conditions [Bürgmann and Dresen, 2008]. The same is true for the viscous behavior of e.g. the Earth's interior: While the upper crust is usually assumed to be in a frictional equilibrium with active faults that might limit the strength [Bürgmann and Dresen, 2008], a pressure dependent increase of frictional strength with depth is counteracted by thermally activated creep processes which reduce the viscous strength with increasing temperature and depth [Goetze & Evans 1979; Goetze, 1978; Hirt and Kohlstedt, 2003; McBirney and Murase, 1984; Melosh, 1980].

A simple linear viscous rheology is not sufficient for a realistic calculation of mantle behavior. We therefore use a more sophisticated model that has been previously presented in several papers (e.g. Kameyama et al., 1999; Kawazoe et al., 2009; Karato and Wu, 1993). This model consists of three different regimes depending on stress and temperature: Grain-size dependent *diffusion creep*, a power-law *dislocation creep* (see also e.g. Faul et al., 2011), and the exponential *Peierls mechanism* for larger stresses. While deformation caused by classic geodynamical mantle convection is dominated by diffusion and dislocation creep, the Peierls mechanism becomes essential for deformation at larger stresses and strain rates, such as during meteorite impact and the subsequent collapse of the initial deep crater.

In our approach the overall strain rate due to inelastic deformation $\varepsilon_v$ consists of a sum of viscous strain rates due to the three creep mechanisms: Diffusion creep ($\dot{\varepsilon}_1$), Dislocation (or power-law) creep ($\dot{\varepsilon}_n$), and Peierls creep ($\dot{\varepsilon}_P$).

$$\dot{\varepsilon}_v = \dot{\varepsilon}_1 + \dot{\varepsilon}_n + \dot{\varepsilon}_P \qquad (1)$$

**Diffusion creep** accounts for deformation of material by diffusion of vacancies in the crystal lattice. It is therefore a plastic deformation, which gives rise to ductile, not brittle, behavior. It is generally a linear function of stress. The strain rate due to diffusion creep depends not only on temperature $T$ and deviatoric stress $\sigma$ (this is usually taken to equal the second invariant of the stress tensor, as the appropriate generalization of shear stress) but also on grain size $a$:

$$\dot{\varepsilon}_1 = \frac{\sigma}{\eta_0} \left(\frac{a}{a_0}\right)^{-m} \cdot exp\left(-\frac{E_1}{R \cdot T}\right) \qquad (2)$$

where $a_0$ is the reference grain size (here: $a_0$=1 mm), $R$ is the gas constant, $\eta_0$ is a reference viscosity, and $E_1$ is an activation enthalpy. $m$ is a material-dependent parameter describing the grain size-dependency of diffusion creep, which ranges between 2 for volume diffusion (Nabarro-Herring creep) and 3 for grain boundary diffusion (Coble creep). Diffusion creep is mainly important at lower stresses, is highly temperature dependent and results in very small strain rates.

***Dislocation creep (power-law creep)*** describes material deformation as the result of dislocations moving through the crystal lattice (see also Carrez and Cordier, 2010). It is best described by a power-law function of the deviatoric stress, in which the strain rate depends on temperature and stress but is independent of the grain size:

$$\dot{\varepsilon}_n = \frac{\sigma_c}{\eta_0}\left(\frac{\sigma}{\sigma_c}\right)^n \cdot exp\left(-\frac{E_n}{R \cdot T}\right) \qquad (3)$$

Here $E_n$ is the activation enthalpy of non-Newtonian creep, $\sigma_c$ is the critical stress, and $n$ is the exponent for the power-law dependence of this creep that usually ranges between 3 and 5. This mechanism usually predominates at intermediate stresses and higher temperatures.

The ***Peierls mechanism*** is a special type of dislocation creep that is thermally activated and dominates only at high deviatoric stress. The creep law for Peierls mechanism is given by

$$\dot{\varepsilon}_P = A_P \cdot exp\left[-\frac{E_P}{R \cdot T}\left(1 - \frac{\sigma}{\sigma_P}\right)^q\right] \qquad (4)$$

where $A_P$ and $q$ are material-dependent parameters, $E_P$ is the activation enthalpy for the Peierls mechanism, and $\sigma_P$ is the Peierls stress. The Peierls creep dominates for stresses above about 500 MPa, although this depends upon the material: Mantle materials such as olivine are much more susceptible to Peierls creep than typical crustal minerals such as plagioclase.

The parameters of the equations shown above are strongly material dependent. Since our study focuses on the mantle rheology we chose values that best reflect dry olivine, as published in *Kameyama et al., 1999*. See Table 1 below for the parameters.

Table 1 Parameters and their values used for the viscous flow calculation (adopted from Kameyama et al., 1999)

| | Description | Parameter value |
|---|---|---|
| $\eta_0$ | Reference viscosity | $3.88 \times 10^{10}$ [Pa s] |
| $a_0$ | Reference grain size | 1 [mm] |
| $R$ | Universal gas constant | 8.31 [J/mol K] |
| $E_1$ | Activation enthalpy for diffusion creep | $3.0 \times 10^5$ [J/mol K] |
| $E_n$ | Activation enthalpy for the power-law creep | $5.4 \times 10^5$ [J/mol K] |
| $E_P$ | Activation enthalpy for the Peierls mechanism | $5.4 \times 10^5$ [J/mol K] |
| $\sigma_c$ | Critical stress | 91.25 [Pa] |
| $\sigma_p$ | Peierls stress | $8.5 \times 10^9$ [Pa] |
| $m$ | Exponent for grain-size dependence (diffusion creep) | 2.5 |
| $n$ | Exponent for stress dependence (power-law creep) | 3.5 |
| $q$ | Exponent for stress dependence (Peierls mechanism) | 2.0 |

Expanding equations (1)-(4), the overall viscous strain rate is given by

$$\dot{\varepsilon}_v = \overbrace{\frac{\sigma}{\eta_0}\left(\frac{a}{a_0}\right)^{-m} exp\left(-\frac{E_1}{RT}\right)}^{\text{diffusion creep}} + \overbrace{\frac{\sigma_c}{\eta_0}\left(\frac{\sigma}{\sigma_c}\right)^n exp\left(-\frac{E_n}{RT}\right)}^{\text{dislocation creep}} + \overbrace{A_P \cdot exp\left[-\frac{E_P}{RT}\left(1-\frac{\sigma}{\sigma_P}\right)^q\right]}^{\text{Peierls creep}} \quad (5)$$

The deformation map in Figure 1 shows the relationship between strain rate, stress, and temperature as given by equation (5). We can clearly see that the *diffusion creep* only accounts for strain rates which are (usually) much smaller than those expected for large-scale planetary impact events, except for much slower viscous relaxation long after the crater forms. Thus, it is likely that diffusion creep does not play an important role for this study, where typical strain rates range from $1\times10^{-3}$ to 10 s$^{-1}$. For performance reasons equation (2) might be neglected if applied for planetary scale impact simulations. In the framework of this study, however, we included this mechanism for completeness. Instead we ignored a possible pressure dependence in the *Peierls regime* (as e.g. suggested in Kawazoe et al., 2009), because it is much smaller than the temperature and strain rate dependence for the simulations intended within the framework of our study. Indeed, if long-term crater relaxation is the goal of a simulation, then iSALE is a poor choice, because the inclusion of inertial accelerations in iSALE requires very small time steps that would be impractical for deformation occurring over millions of years.

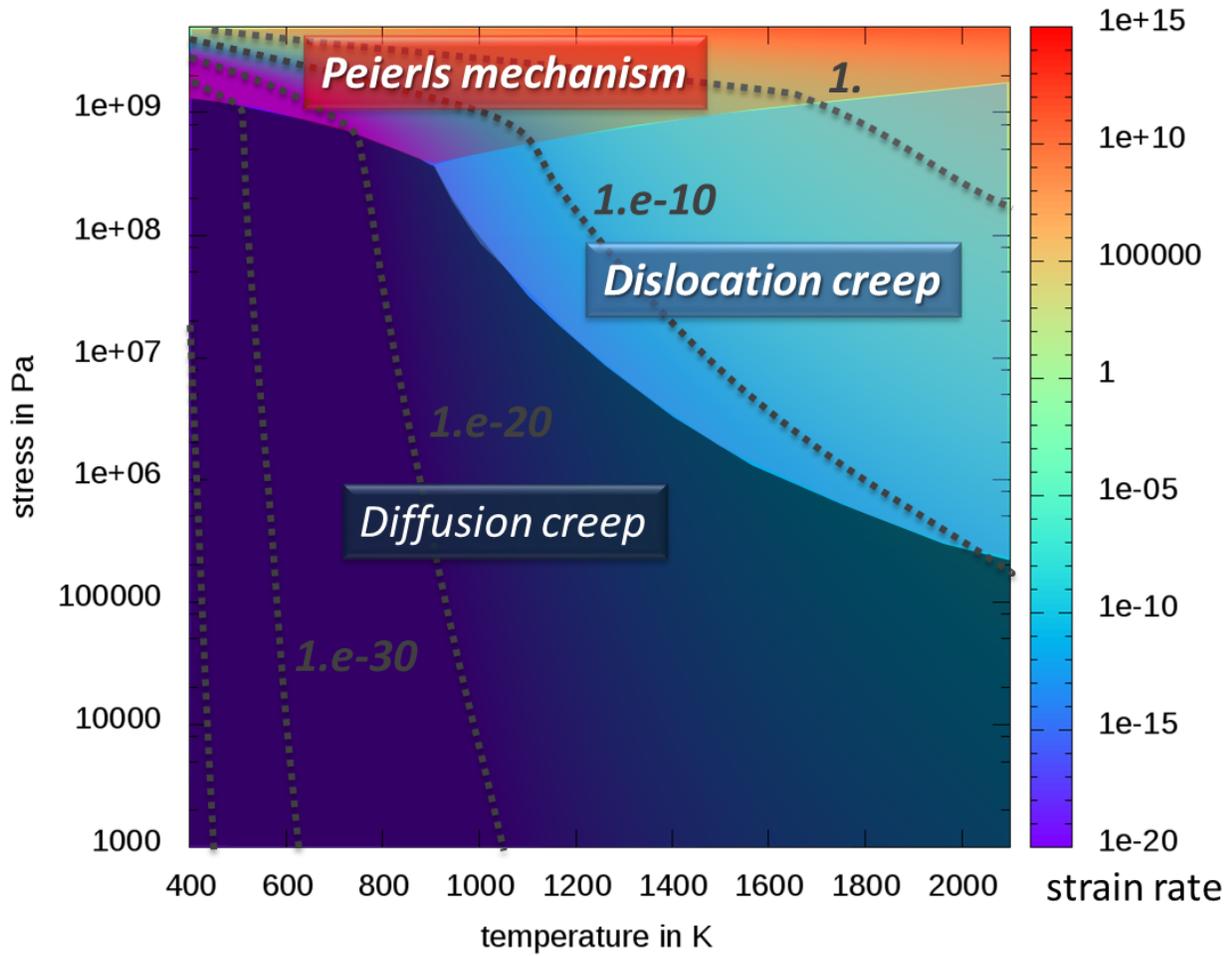

Figure 1 Strain rate as a function of stress and temperature according to Eq. (5). The parameter space dominated by each of the terms in eq. (5) is denoted by different hues, depending on strain rate. Note that the original equations implicitly include a pressure dependence in the enthalpy term, which is not strong and therefore neglected here. The *diffusion regime* is highly temperature dependent and important for small strain rates only. *Dislocation creep* occurs for intermediate stresses and higher temperatures, while the *Peierls mechanism* dominates at higher stresses (>500 MPa) and shows astrong stress dependence for low temperatures. The dashed contours are strain rate in units of sec$^{-1}$.

## 3 Coupling the Peierls mechanism with an elasto-plastic model

The time dependence of the response of a viscoelastic system is similar to that of an electrical circuit. From a mathematical perspective both systems can be described by an identical set of ordinary differential equations. A convenient approach to this problem is the so-called "spring-dashpot" models, where the elastic rheology is represented by an elastic spring, described by

$$\sigma = k \cdot \varepsilon \qquad (6)$$

Here $\sigma$ and $\varepsilon$ represent the spring force and displacement, and the spring constant $k$ plays the role of the shear modulus $\mu$. The viscous rheology can be visualized by a Newtonian dashpot

$$\sigma = \eta \cdot \dot{\varepsilon} \qquad (7)$$

where η is the viscosity. These two basic elements can be combined in different ways to simulate more complex rheologic behavior.

Following this approach, different rheologic models are in general use: The *Maxwell-configuration* connects one spring and one dashpot in series, so that the strain is the sum of each element while the stress is the same in both. The Kelvin-Voigt puts a viscous and an elastic element in parallel so that the strain is equal in both elements while the total stress is the sum of the individual stresses. A *Standard Linear Solid* combines a Maxwell element and a spring in parallel. To describe a more complex rheology such as that of polymers, mathematical solutions have been presented that rely upon arbitrary combinations of multiple elements [Wiechert, 1893; Tschoegl, 1989; Brinson and Brinson, 2008].

We used a *Maxwell-configuration* in conjunction with a plastic (*Bingham*) flow model to connect the viscous creep rheology as described above to an elasto-plastic model to allow visco-elastic or even visco-elasto-plastic material behavior. When using elasticity, viscosity, and plasticity in series, every element in the entire system is experiences the same stress while the strain rates are additive:

$$\sigma_T = \sigma_v = \sigma_e = \sigma_p \tag{7}$$

$$\dot{\varepsilon}_T = \dot{\varepsilon}_v + \dot{\varepsilon}_e + \dot{\varepsilon}_p \tag{8}$$

The subscripts *T*, *v*, *e*, and *p* refer to the total, viscous, elastic, and plastic contributions, respectively. The constitutive equation for viscoelasticity in a Maxwell-configuration is directly inherited from equations (6), (7), and (8):

$$\dot{\varepsilon} = \frac{1}{\mu}\dot{\sigma} + \frac{1}{\eta}\sigma \tag{9}$$

The advantage of utilizing a Maxwell-model is that is permits relaxation of stresses, while allowing arbitrarily large strains, consistent with observations of the behavior of geologic materials. For example, under the condition of a strain that is suddenly applied and then held fixed, the stress relaxes by

$$\sigma(t) = \sigma_0 \cdot e^{-\frac{t}{T_M}} \tag{10}$$

where $T_M = \frac{\eta}{\mu}$ is the Maxwell decay time.

## 4  Implementation of visco-elasto-plasticity in a hydrocode

The constitutive equation for viscous creep, equation (5), shows how to calculate the strain rate from a given stress. However, the hydrocode we utilize for our study, iSALE, works in the opposite way: it updates the stresses from a given strain rate. This creates two complications that must be solved:

1. The constitutive equation (5) must be inverted to compute the stress from a given strain rate. Because this equation is a transcendental equation it must be solved numerically, which is very time-consuming.
2. Because we calculate the stress contributions from the total strain rates and we assume that each element acts in series, we need to determine the correct partitioning of the total strain rate ($\dot{\varepsilon}_T$) into an elastic strain rate ($\dot{\varepsilon}_e$), a viscous strain rate ($\dot{\varepsilon}_v$) and a plastic strain rate ($\dot{\varepsilon}_p$).

   In the series configuration the same stress acts on each element, which responds with its own strain rate, according to the size of each term in equation (5). Please note: when we use *deviatoric* stress and strain, these equations hold separately for each component of the stress and strain.

In the following, we present an approach to deal with these issues. This approach to the computation of visco-elasto-plastic behavior requires four different steps which are described in this section:

1. Derive an effective viscosity from the previous stress.
2. Calculate an effective stress from the effective viscosity.
3. Use the effective stress to calculate the corresponding viscous and elastic strain rates.
4. Reduce the stress and update strain rates if the stress exceeds the Bingham yield envelope (plastic deformation).

## 4.1 Derive an effective viscosity

For viscous material behavior, we consider diffusion creep, dislocation creep, and Peierls creep. The strain rate due to each mechanism is additive, which results in the main equation (5). Note that most experimenters use pure shear, so they measure $2 \cdot \varepsilon_{ij}$ and not $\varepsilon_{ij}$. Thus, our basic equation is

$$2 \cdot \dot{\varepsilon} = \overbrace{\left(\frac{a}{a_0}\right)^{-m} \frac{\sigma}{\eta_0} e^{-\frac{E_1}{RT}}}^{\text{diffusion creep}} + \overbrace{\frac{\sigma_c}{\eta_0}\left(\frac{\sigma}{\sigma_c}\right)^n e^{-\frac{E_n}{RT}}}^{\text{dislocation creep}} + \overbrace{A_P \cdot e^{-\frac{E_P}{RT}\left(1-\frac{\sigma}{\sigma_P}\right)^2}}^{\text{Peierls creep}}$$

$$= \left[\frac{1}{\eta_0}\left(\frac{a}{a_0}\right)^{-m} e^{-\frac{E_1}{RT}} + \frac{1}{\eta_0}\left(\frac{\sigma}{\sigma_c}\right)^{n-1} e^{-\frac{E_n}{RT}} + \frac{A_P}{\sigma} \cdot e^{-\frac{E_P}{RT}\left(1-\frac{\sigma}{\sigma_P}\right)^2}\right] \sigma$$

(11)

Considering the constitutive equation for the viscous element

$$\dot{\varepsilon}_{ij'} = \frac{1}{2\eta}\sigma_{ij'}$$

(12)

we can easily compute $\eta_{\text{eff}}$ at any working stress σ:

$$\frac{1}{\eta_{\text{eff}}} = \frac{1}{\eta_0}\left(\frac{a}{a_0}\right)^{-m} e^{-\frac{E_1}{RT}} + \frac{1}{\eta_0}\left(\frac{\sigma}{\sigma_c}\right)^{n-1} e^{-\frac{E_n}{RT}} + \frac{A_P}{\sigma} \cdot e^{-\frac{E_P}{RT}\left(1-\frac{\sigma}{\sigma_P}\right)^2}$$

Here we use the second invariant of the old stress tensor as a reference stress.

## 4.2 Calculate the effective stress

At the beginning of each time step we know the previous stress $\sigma_{ij}(t_0)$. In our approach viscosity only contributes to the stress of "intact" material, i.e. material whose stress does not exceed the yield envelope ($\sigma_{II}$<Y). In this case, no plastic strain occurs ($\dot{\varepsilon}_p = 0$). For the current (i.e. new) time step, we therefore know

$$\dot{\varepsilon}_T = \dot{\varepsilon}_e + \dot{\varepsilon}_v \quad (\dot{\varepsilon}_p = 0) \tag{13}$$

Using the constitutive equations (6) and (7) we obtain

$$\underbrace{\dot{\varepsilon}_T}_{known} = \frac{1}{2\mu}\dot{\sigma} + \frac{1}{2\eta}\sigma \tag{14}$$

Please note that here we use the effective viscosity $\eta = \eta_{eff}$. The differential equation for σ is given by

$$2\mu\dot{\varepsilon} = \frac{d\sigma}{dt} + \frac{\mu}{\eta}\sigma = \frac{d\sigma}{dt} + \frac{\sigma}{T_M} \tag{15}$$

where $\frac{\eta}{\mu} = T_M$ is the *Maxwell decay time*. This results in the general solution given by

$$\sigma(t) = \sigma(t_0)e^{-\frac{t-t_0}{T_M}} + e^{-\frac{t}{T_M}} \cdot \int_{t_0}^{t} \dot{\varepsilon}_T(t) \cdot 2\mu \cdot e^{\frac{t'}{T_M}} dt' \tag{16}$$

If $t - t_0 = \Delta t$ is short, and we assume that $\dot{\varepsilon}_T(t)$ is constant over $t_0$ to $t_0 + \Delta t$, then

$$\sigma(t) = \sigma(t_0)e^{-\frac{\overbrace{t}^{=t_0+\Delta t}}{T_M}} + 2\mu\dot{\varepsilon}(t_0)\left[1 - e^{-\frac{\Delta t}{T_M}}\right] \cdot T_M \tag{17}$$

$$\sigma(t) \approx \sigma(t_0)e^{-\frac{\Delta t}{T_M}} + 2\mu\dot{\varepsilon}\Delta t \tag{17}$$

$$\sigma(t) \approx \sigma(t_0)\left(1 - \frac{\Delta t}{T_M}\right) + 2\mu\dot{\varepsilon}\Delta t \tag{17}$$

to first order in $\Delta t$. Now we can use equation (17) to calculate the effective stress from the effective viscosity.

## 4.3 Partitioning viscous and elastic strain rates

By using the derived effective stress, we can now compute the elastic and viscous strain easily. The elastic strain rate is given by

$$\dot{\varepsilon}_e = \frac{1}{2\mu}\dot{\sigma} = \frac{1}{2\mu}\left[-\sigma(t_0)\frac{\Delta t}{T_M} + 2\mu\Delta t \dot{\varepsilon}_T\right]\frac{1}{\Delta t} \qquad (18)$$

$$\dot{\varepsilon}_e = -\underbrace{\frac{1}{2\eta}\sigma(t_0)}_{\text{viscous contrib.}} + \underbrace{\dot{\varepsilon}_T}_{\text{new strain in this cycle}} \qquad (19)$$

and the viscous strain rate is obtained by

$$\dot{\varepsilon}_v = \frac{1}{2\eta}\sigma = \frac{1}{2\eta}\sigma(t_0)\left(1 - \frac{\Delta t}{T_M}\right) + \frac{\dot{\varepsilon}}{T_M}\Delta t \qquad (20)$$

## 4.4 Plasticity – rheology above the yield envelope

Finally, we compare the new stress (here: the second invariant $\sigma_{II}$) with the yield envelope. If $\sigma_{II} <$ Y, no plastic strain occurs and the total strain rate is partitioned into elastic and viscous strain rates as given in the equations (18) and (20) above. If the stress exceeds the yield limit ($\sigma_{II} \geq Y$), then stresses relax to the yield envelope

$$\sigma_{ij}^{new} = \frac{Y}{\sigma_{II}}\sigma_{ij}(t_0) \qquad (21)$$

and, once the new stress is known, we can use again the constitutive equations to compute the new (reduced) elastic and viscous strain rates $\dot{\varepsilon}_e$ and $\dot{\varepsilon}_v$ at a stress on the yield envelope:

$$\varepsilon_e = \frac{1}{2\mu}\sigma_{ij}^{new} \qquad (22)$$

$$\dot{\varepsilon}_v = \frac{1}{2\eta}\sigma_{ij}^{new} \qquad (23)$$

The remaining part contributes to the plastic strain rate

$$\dot{\varepsilon}_p = \dot{\varepsilon}_T - \dot{\varepsilon}_e - \dot{\varepsilon}_v \qquad (24)$$

# 5 Results and validation

To test and explore our rheology model we performed numerical simulations of a 1 m sized block sheared with a velocity of 1 m/s and recorded the resulting stresses and effective viscosities. While this test becomes quite unrealistic after strains exceed a few tens of percent, it provides a stringent test of the stability of our numerical method. The block is composed of dunite. To calculate the thermodynamic behavior we used the *Analytical Equation of State* (ANEOS; see Thompson and Lauson, 1972) with tabularized data for dunite. For the viscous calculations we used the same parameters as those listed in Table 1. We assumed a pre-heated block at an initial temperature of T=1700 K to focus on the regime where the Peierls creep becomes important (see also Figure 1). This temperature also better reflects the thermal conditions in the (Earth) mantle.

The elastic stress contribution $\sigma_e$ is calculated according to equation (17) by

$$\sigma_e = 2\mu\dot{\varepsilon}\Delta t \quad (25)$$

where the elastic shear modulus $\mu$ is computed from the speed of shear waves and density and, thus, on the thermodynamic behavior of the material. At the conditions described above, it is on the order of 130 GPa.

The plastic behavior is simulated by limiting stresses to the yield envelope and reducing strain rates properly (see section 4.4). The iSALE hydrocode contains different methods for calculating the yield envelope of a given material. For these tests we utilized the so-called rock-model (see Collins et al., 2004). In this model, the yield envelope comprises different stress paths for intact and damaged material. While the yield envelope for completely damaged material is calculated by using a simple Mohr-Coulomb approach, intact material is represented by a Lundborg relationship. Between these two states a linear interpolation is used to retrieve an adequate yield limit. The parameters used for this plasticity model are listed in Table 2.

**Table 2 Parameters used to describe the elasto-plastic and thermal behavior of dunite.**

| Parameter | |
|---|---|
| Poisson ratio | 0.3 |
| Equation of state | ANEOS for dunite |
| Specific heat capacity | 1000. J kg$^{-1}$ K$^{-1}$ |
| Shear strength at initial condition (intact material) | 10 MPa |
| Coefficient of internal friction (intact material) | 1.1 |
| Shear limit (intact material) | 2.5 GPa |
| Shear strength at initial condition (damaged material) | 10 KPa |
| Coefficient of internal friction (damaged material) | 0.8 |
| Shear limit (damaged material) | 2. GPa |

We calculated the shearing of the block for different material rheologies: (a) purely elastic, (b) elasto-plastic, (c) visco-elastic and (d) visco-elasto-plastic. The results are shown in Figure 2ff. Stresses and strain rates are always represented by their XY-component. Figure 2 shows the time evolution of stress for all four different rheologies. While a purely elastic rheology results in a significant and continuous stress increase, stresses in the elasto-plastic regime are significantly lower. In the latter case the resulting stresses are always above the yield envelope. Thus, when considering elasto-plastic behavior, plastic work and plastic strain occurs at any time during deformation.

A comparison between the elastic and visco-elastic behavior reveals an identical stress evolution in the very early stage. Later on, however, stresses begin to relax in the viscoelastic case. The rate of the stress decay is defined by the Maxwell time $T_M = \eta/\mu$ which describes the time required to decrease the stress to 1/e of the initial stress. In laboratory experiments this measure is often used to obtain insights into the viscous behavior of a given material. In our approach, however, the Maxwell time is itself time-dependent, because the effective viscosity $\eta$ varies with time, as shown in Figure 3. In this experiment, the Maxwell time quickly declines from 35 s in the beginning to less than 30 ms after 0.1 s from the beginning of the shearing process. In the same time the effective viscosity decreases over three orders of magnitude from > $10^{12}$ Pa-s to 1 GPa-s.

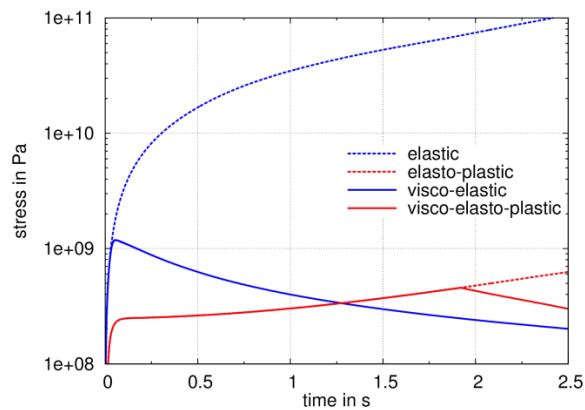

**Figure 2** Evolution of stress during shearing for elastic, elasto-plastic, visco-elastic and visco-elasto-plastic material behavior. The rollover of the elastic stresses after about 0.1 s is due to the extreme distortion of the initial 1 x 1 x 1 m cube of dunite (the strain is 10% at 0.1 s). At 1 s the total strain is 100% and continues to increase thereafter as the block is distorted from a cube into a long parallelepiped. Our algorithm nevertheless faithfully follows the stress evolution in this distorted block. Note that in our rheological model the stress is the same in each element of the system, so the stress plotted acts equally on the elastic, viscous and plastic elements.

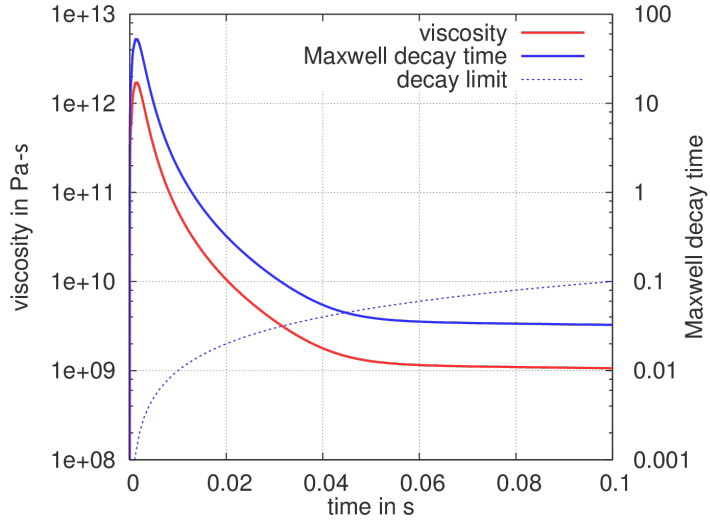

**Figure 3** Effective viscosity and corresponding Maxwell time for viscoelastic material behavior during shearing (T=1700 K). The function f(x) = x is plotted as a dashed line to illustrate whether the current time step is beyond the corresponding Maxwell time or not.

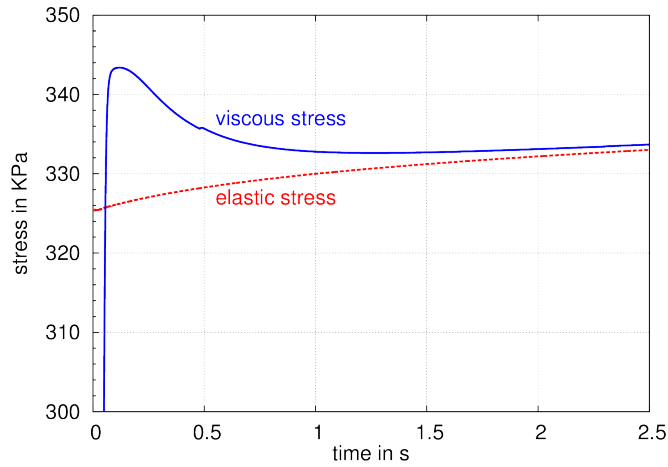

**Figure 4** Viscous and elastic stress contributions during shearing (T=1700 K). Note that the magnitude of the contributions depends on the time increment $\Delta t$, as in Equations (29) through (31).

Figure 4 shows the elastic ($\sigma_e$) and the viscous ($\sigma_v$) contributions to the stress:

$$\sigma_e = 2\mu\dot{\varepsilon}\Delta t \tag{26}$$

$$\sigma_v = \sigma(t_0)\frac{\Delta t}{T_M} = \sigma(t_0)\frac{\mu\Delta t}{\eta} \tag{27}$$

$$\sigma = \sigma(t_0) + \sigma_e - \sigma_v \tag{28}$$

Please note that the magnitude of these contributions is strongly dependent on the size of the time increment ($\Delta t$), as equations (17) and (28) illustrate. Also note the negative sign in the viscous term of this equation indicating that the viscous contribution counteracts the elastic stresses. As is also evident in Figure 3, the rheology is initially dominated by elastic (or elasto-plastic) behavior. The viscous stresses are significantly lower than the elastic stresses, but increase rapidly. Thus, very shortly after the onset of shearing, viscous stresses become more prominent than the elastic contributions. The stress therefore decays. This happens roughly when the Maxwell time and the current time step are the same (at the intersection between the Maxwell decay time and the decay limit in Figure 3). The dominance of the viscous contributions results in a quick decay of the total stresses. While the elastic stresses increase slowly, the viscous stresses decrease after reaching their maximum. Thus, the decrease of stress is also damped by time, in agreement with equation (10).

When considering a full visco-elasto-plastic rheology, the material initially behaves as purely elasto-plastic (see Figure 2) because there is not time for viscous strains to accumulate. The resulting stresses in the visco-elastic and visco-elasto-plastic scenarios are exactly the same until the block is strongly deformed and the Bingham yield stress is exceeded. Despite stress relaxation in the viscous element (and the corresponding accumulation of viscous strain), the resulting stresses are above the yield envelope during large parts of the deformation. Stresses are, thus, reduced (see section 4.4) and part of both the viscous and elastic strain is converted into plastic strain. At some later time, however, the viscous flow relaxes the stress below the yield envelope. From this moment on, the resulting stress veers away from the yield envelope and the material behaves as purely visco-elastic.

All three regimes considered by us to calculate a realistic viscous behavior – Diffusion creep, dislocation creep and the Peierls mechanism –depend on temperature. Each of these regimes is activated or dominant at different temperature conditions, as Figure 1 illustrates. Thus, we repeated the experiments described above and systematically varied the initial temperature of the block material. Figure 5 shows the stress evolution for both purely elastic (dashed lines) and visco-elastic (solid lines) material behavior and for different temperatures. It is evident that the resulting stresses decrease with increasing temperatures. If only the viscous behavior is considered, however, the stresses are much lower and even the temperature dependence on stress is much more pronounced. Furthermore, also the separation of the visco-elastic path from the purely elastic path occurs much earlier for higher temperatures, as becoming visible in Figure 5 (right), which shows the stress evolution for the very early stage. This observation is also demonstrated in Figure 6, showing the effective viscosity as a function of time. The initial viscosity is much higher for lower temperatures. At room temperature, the resulting initial viscosity $\eta(t \approx 0)$ is even above $10^{60}$ Pa-s and the material behaves therefore more solid-like until viscosity starts to decrease after some time due to the stress dependence of the viscosity. The onset of a significant and nonlinear decrease of the viscosity depends strongly on temperature and occurs much earlier for higher temperatures. At a temperature of 2000 K the initial viscosity drops down to 10 GPa-s and viscous effects are therefore more prominent right from the beginning of the deformation. Despite the different temperature conditions all viscosities asymptotically approach the same value at a later stage (here ≈10 GPa s). The resulting stresses for T=2000 K are a bit lower which can be explained by the onset of changes in the thermodynamic behavior (beginning phase transitions) at conditions close to the melting point of dunite.

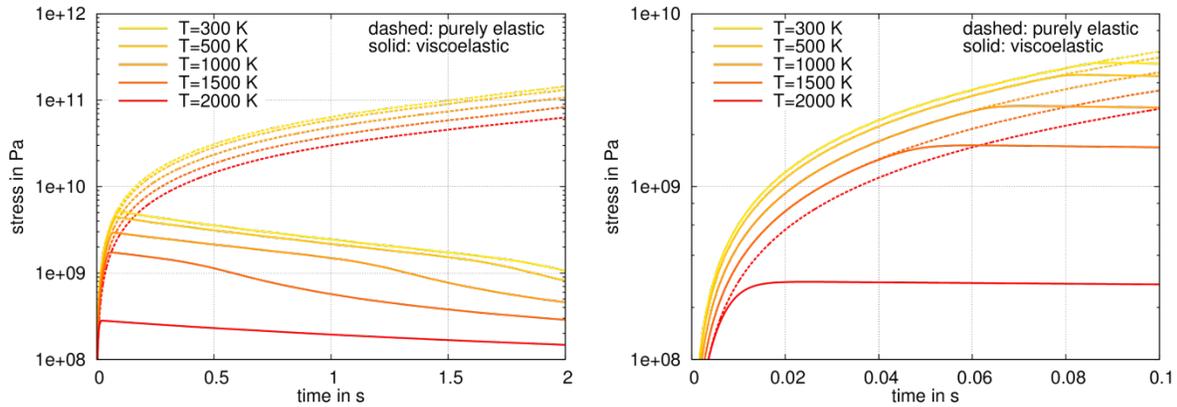

Figure 5 Temperature dependence of elastic and viscoelastic material rheology. The right panel shows an expanded section of the initial stage.

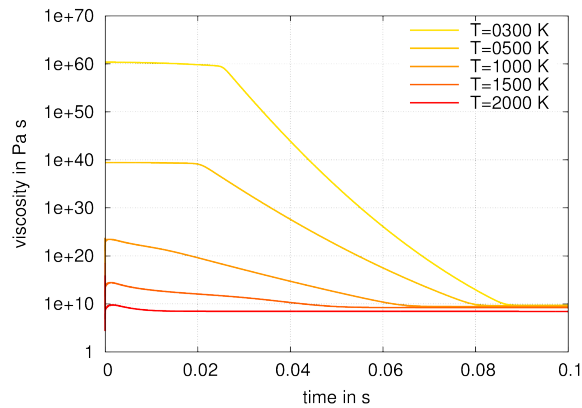

Figure 6 Dependence of the evolution of effective viscosity during shearing on temperature.

# 6   Summary and conclusion

We have presented a simple and efficient way of implementing a viscous creep rheology based on Diffusion creep, Dislocation creep and the Peierls mechanism in conjunction with an elasto-plastic rheology model into a shock-physics code. The three regimes are combined continuously, i.e. there is no threshold initiating a sudden change from one regime to the next. While the viscoelasticity of the mantle as utilized for mantle convection or other slow geodynamic or geological processes is mostly controlled by diffusion and dislocation creep, we showed that the Peierls mechanism becomes the dominant mechanism at the large stresses and strain rates occurring during meteorite impacts.

Our approach to implementing the model into the shock-physics-code iSALE is based on the calculation of an "effective viscosity" which is then used as a reference viscosity for any underlying viscoelastic (or even visco-elasto-plastic) model. Here we use a Maxwell-configuration, where elastic, viscous and plastic strain rates sum up to the total strain and the stress contributions at the viscous, elastic and plastic elements are identical). The *Maxwell-model* best describes stress relaxation (see equation (10)) in natural materials, and so is likely to be important for the formation of large meteorite impact basins, for which this code was intended. However, this model does not consider the full range of observed creep phenomena accurately. Under constant stress conditions, strain in this model increases linearly with time. Most natural materials, however, exhibit a more complex relationship and at lower temperatures strain rates usually decrease with time, in a regime known as primary creep. The *Kelvin-Voigt* approach, where the viscous and elastic elements are arranged in parallel (and, thus, each element encounters the same strain rate, but the resulting stresses sum up to the total stress), simulates creep at small strains much more accurately, but it is less accurate for predicting large strain behavior. In the last decades more complex rheologies have been developed and proposed, such as *Standard-Linear-Solid* (where a Maxwell-model is combined with an additional elastic element in parallel) or a *Generalized Maxwell Model* or *Maxwell-Wiechert model* (arbitrary number of Maxwell-elements in parallel, see Wiechert, 1893; Tschoegl, 1989; Brinson and Brinson, 2008). Our approach of calculating an effective viscosity can be easily applied to these more complex configurations, which we intend to do in future work.

# 7 Acknowledgement

This work is funded by NASA's GRAIL mission and the Helmholtz-Alliance HA-203, "Planetary Evolution and Life" by the Helmholtz-Gemeinschaft Deutscher Forschungszentren (HGF)

# References

Benson, D. J. (1991). A new two-dimensional flux-limited shock viscosity for impact calculations. Computer Methods in Applied Mechanics and Engineering, 93(1), 39-95.

Brinson, H.F., & Brinson, L.C. (2008) Polymer Engineering Science and Viscoelasticity - An Introduction, Springer, Berlin.

Bürgmann, R. and Dresen, G. (2008) Rheology of the Lower Crust and Upper Mantle: Evidence from Rock Mechanics, Geodesy, and Field Observations. Annu. Rev. Earth Planet. Sci. 36: 531-567

Carrez, P., & Cordier, P. (2010). Modeling dislocations and plasticity of deep earth materials. Reviews in Mineralogy and Geochemistry, 71(1), 225-251.

Christeson, G. L., Collins, G. S., Morgan, J. V., Gulick, S. P., Barton, P. J., & Warner, M. R. (2009). Mantle deformation beneath the Chicxulub impact crater. Earth and Planetary Science Letters, 284(1), 249-257.

Collins, G. S., Melosh, H. J., and Ivanov, B. A. (2004). Modeling damage and deformation in impact simulations. Meteoritics & Planetary Science, 39, pp. 217–231.

Collins, G. S. (2014) Numerical simulations of impact crater formation with dilatancy. J. Geophys. Res. - Planets, 10.1002/2014JE4708 pp. 1-20.

Darabi, M. K., Abu Al-Rub, R. K., Masad, E. A., Huang, C. W., & Little, D. N. (2011). A thermo-viscoelastic–viscoplastic–viscodamage constitutive model for asphaltic materials. International Journal of Solids and Structures, 48(1), 191-207.

Elbeshausen D. and Wünnemann K. (2011). iSALE-3D: A three-dimensional, multi-material, multi-rheology hydrocode and its applications to large-scale geodynamic processes. In: Schäfer, F. and Hiermaier, S. (editors): Proceedings of the 11th Hypervelocity Impact Symposium 2010, Freiburg,

Germany, April 11-15, 2010. Fraunhofer Verlag, Stuttgart, 828 pp. (Epsilon. Forschungsergebnisse aus der Kurzzeitdynamik, 20) (ISBN 3-8396-0280-7 ; ISBN 978-3-8396-0280-5 ; ISSN 1612-6718)

Elbeshausen, D., Wünnemann, K., and Collins, G. S. (2009). Scaling of oblique impacts in frictional targets: Implications for crater size and formation mechanisms. Icarus, 204 (2), pp. 716–731.

Faul, U. H., Fitz Gerald, J. D., Farla, R. J. M., Ahlefeldt, R., & Jackson, I. (2011). Dislocation creep of fine-grained olivine. Journal of Geophysical Research: Solid Earth (1978–2012), 116(B1).

Gennisson, J. L., Deffieux, T., Macé, E., Montaldo, G., Fink, M., & Tanter, M. (2010). Viscoelastic and anisotropic mechanical properties of in vivo muscle tissue assessed by supersonic shear imaging. Ultrasound in medicine & biology, 36(5), 789-801.

Goetze, C. (1978). The Mechanisms of Creep in Olivine. Royal Society of London Philosophical Transactions Series A, 288, 99-119.

Goetze, C., & Evans, B. (1979). Stress and temperature in the bending lithosphere as constrained by experimental rock mechanics. Geophysical Journal International, 59(3), 463-478.

Grooman, B., Fujiwara, I., Otey, C., & Upadhyaya, A. (2012). Morphology and Viscoelasticity of Actin Networks Formed with the Mutually Interacting Crosslinkers: Palladin and Alpha-actinin. PloS one, 7(8), e42773.

Hirt, C.W., Amsden, A.A. and Cook, J.L. (1974). An arbitrary Lagrangian-Eulerian computing method for all flow speeds. Journal of Computational Physics, 14(3), pp. 227-253.

Hirth, G., & Kohlstedt, D. (2003). Rheology of the upper mantle and the mantle wedge: A view from the experimentalists. Geophysical Monograph Series, 138, 83-105.

Hobbs, B., Regenauer-Lieb, K., & Ord, A. (2007). Thermodynamics of folding in the middle to lower crust. Geology, 35(2), 175-178.

Hong, W. (2011). Modeling viscoelastic dielectrics. Journal of the Mechanics and Physics of Solids, 59(3), 637-650.

Ivanov, B. A. (2005). Numerical modeling of the largest terrestrial meteorite craters. Solar System Research, 39(5), 381-409.

Ivanov, B. A., Melosh, H. J., & Pierazzo, E. (2010). Basin-forming impacts: Reconnaissance modeling. Geological Society of America Special Papers, 465, 29-49.

Ivanov, B. A., Deniem, D., and Neukum, G. (1997). Implementation of dynamic strength models into 2D hydrocodes: Applications for atmospheric breakup and impact cratering. International Journal of Impact Engineering, 20:411--430.

Kameyama, M., Yuen, D. A., & Karato, S. I. (1999). Thermal-mechanical effects of low-temperature plasticity (the Peierls mechanism) on the deformation of a viscoelastic shear zone. Earth and Planetary Science Letters, 168(1), 159-172.

Karato, S. I. (2010). Rheology of the Earth's mantle: A historical review. Gondwana Research, 18(1), 17-45.

Karato, S. I., & Wu, P. (1993). Rheology of the upper mantle: A synthesis. Science, 260(5109), 771-778.

Kawazoe, T., Karato, S. I., Otsuka, K., Jing, Z., & Mookherjee, M. (2009). Shear deformation of dry polycrystalline olivine under deep upper mantle conditions using a rotational Drickamer apparatus (RDA). Physics of the Earth and Planetary Interiors, 174(1), 128-137.

Lange, H., Casassa, G., Ivins, E. R., Schröder, L., Fritsche, M., Richter, A., Groh, A., & Dietrich, R. (2014). Observed crustal uplift near the Southern Patagonian Icefield constrains improved viscoelastic Earth Models. Geophysical Research Letters 41(3), 805-812.

Larsen, R. J., Dickey, M. D., Whitesides, G. M., and Weitz, D. A. (2009). Viscoelastic properties of oxide-coated liquid metals. Journal of Rheology 53: 1305-1326.

Maxwell, J.C. (1867). On the dynamical theory of gases. Philosophical Transactions of the Royal Society of London 157, 49–88.

McBirney, A. R., & Murase, T. (1984). Rheological properties of magmas. Annual Review of Earth and Planetary Sciences, 12, 337.

Melosh, H. J. (2003). Shock viscosity and rise time of explosion waves in geologic media. Journal of applied physics, 94(7), 4320-4325.

Melosh, H. J. (1989). Impact cratering: A geologic process. Oxford University Press (Oxford Monographs on Geology and Geophysics, No. 11), New York, 253 p.

Melosh, H. J. (1980). Rheology of the Earth: Theory and observation. Physics of the Earth's Interior, 318-336.

Morian, N. E., Alavi, M. Z., Hajj, E. Y., & Sebaaly, P. E. (2014). Evolution of Thermo-Viscoelastic Properties of Asphalt Mixtures with Oxidative Aging. In Transportation Research Board 93rd Annual Meeting (No. 14-1363).

Nabarro, F.R.N. (1948). Deformation of crystals by the motion of single ions, Report of a Conference on Strength of Solids. The Physical Society of London 75–90.

Nguyen, S. T., Jeannin, L., Dormieux, L., & Renard, F. (2013). Fracturing of viscoelastic geomaterials and application to sedimentary layered rocks. Mechanics Research Communications, 49, 50-56.

Nield, G. A., Barletta, V. R., Bordoni, A., King, M. A., Whitehouse, P. L., Clarke, P. J., Domack, E., Scambos, T. A. & Berthier, E. (2014). Rapid bedrock uplift in the Antarctic Peninsula explained by viscoelastic response to recent ice unloading. Earth and Planetary Science Letters, 397, 32-41.

Ohnaka M (1995) A shear failure strength law of rock in the brittle-plastic transition regime. Geophysical Research Letters 22: 25-28

Patton, R. L., & Watkinson, A. J. (2010). Shear localization in solids: insights for mountain building processes from a frame-indifferent ideal material model. Geological Society, London, Special Publications, 335(1), 739-766.

Potter, R. W., Kring, D. A., Collins, G. S., Kiefer, W. S., & McGovern, P. J. (2013). Numerical modeling of the formation and structure of the Orientale impact basin. Journal of Geophysical Research: Planets, 118(5), 963-979.

Remus, F., Mathis, S., Zahn, J. P., & Lainey, V. (2011). The equilibrium tide in viscoelastic parts of planets. In SF2A-2011: Proceedings of the Annual meeting of the French Society of Astronomy and Astrophysics.

Riedel, M. R., & Karato, S. I. (1997). Grain-size evolution in subducted oceanic lithosphere associated with the olivine-spinel transformation and its effects on rheology. Earth and Planetary Science Letters, 148(1), 27-43.

Rouse Jr, P. E. (2004). A theory of the linear viscoelastic properties of dilute solutions of coiling polymers. The Journal of Chemical Physics, 21(7), 1272-1280.

Streitberger, K. J., Wiener, E., Hoffmann, J., Freimann, F. B., Klatt, D., Braun, J., Lin, K., McLaughlin, J., Sprung, C., Klingebiel, R. & Sack, I. (2011). In vivo viscoelastic properties of the brain in normal pressure hydrocephalus. NMR in Biomedicine, 24(4), 385-392.

Swegle, J. W., & Grady, D. E. (1985). Shock viscosity and the prediction of shock wave rise times. Journal of applied physics, 58(2), 692-701.

Teran, J., Fauci, L., & Shelley, M. (2010). Viscoelastic fluid response can increase the speed and efficiency of a free swimmer. Physical review letters, 104(3), 038101.

Theofanous, T. G. (2011). Aerobreakup of Newtonian and viscoelastic liquids. Annual Review of Fluid Mechanics, 43, 661-690.

Thompson, S. and Lauson, H. (1972). Improvements in the CHART D radiation-hydrodynamic code III: revised analytic equations of state. Sandia National Laboratory Report, SC-RR-71 0714:113p.

Tillotson, J. H. (1962). Metallic equations of state for hypervelocity impact. Technical Report GA-3216, General Atomic Report.

Tschoegl N. (1989) The Phenomenological Theory of Linear Viscoelastic Behavior. Springer, Berlin.

Wang, K., Hu, Y., & He, J. (2012). Deformation cycles of subduction earthquakes in a viscoelastic Earth. Nature, 484(7394), 327-332.

Weertman, J. (1955). Theory of steady state creep based on dislocation climb. Journal of Applied Physics 26, 1213–1217.

Wiechert, E. (1893). Gesetze der elastischen Nachwirkung für constante Temperatur. Annalen der Physik, 286, 335–348, 546–570.

Wünnemann K. and Ivanov B. A. (2003): Numerical modelling of impact crater depth-diameter dependence in an acoustically fluidized target. Planetary and Space Science, 51, 831-845.

Wünnemann, K., Collins, G., and Melosh, H. (2006). A strain-based porosity model for use in hydrocode simulations of impacts and implications for transient crater growth in porous targets. Icarus, 180, pp. 514-527.